# Characteristics of alpha projectile fragments emission in interaction of nuclei with emulsion

M. K. Singh[1, 2], Ramji Pathak[1] and V. Singh[2*]

1. Department of physics, Tilakdhari Postgraduate College, Jaunpur - 222002

2. Nuclear and Astroparticle Physics Laboratory

Department of Physics, Banaras Hindu University Varanasi - 221005

The properties of the relativistic alpha fragments produced in interactions of $^{84}$Kr at around 1 A GeV in nuclear emulsion are investigated. The experimental results are compared with the similar results obtained from various projectiles with emulsion interactions at different energies. The total, partial nuclear cross-sections and production rates of alpha fragmentation channels in relativistic nucleus-nucleus collisions and their dependence on the mass number and initial energy of the incident projectile nucleus are investigated. The yields of multiple alpha fragments emitted from the interactions of projectile nuclei with the nuclei of light, medium and heavy target groups of emulsion-detector are discussed and they indicate that the projectile-breakup mechanism seems to be free from the target mass number. It is found that the multiplicity distributions of alpha fragments are well described by the Koba–Nielsen–Olesen (KNO) scaling presentation. The mean multiplicities of the freshly produced newly created charged secondary particles, normally known as shower and secondary particles associated with target in the events where the emission of alpha fragments were accompanied by heavy projectile fragments having Z $\geq$ 3 seem to be constant as the alpha fragments multiplicity increases, and exhibit a behavior independent of the alpha fragments multiplicity.

**PACS numbers:** 29.40.Rg, 25.75.-q, 25.70.Mn,

# I. INTRODUCTION

The projectile fragmentation is relatively well isolated process in the complex scheme of high-energy heavy-ion reactions with multi-baryon system. In the relativistic nucleus-nucleus collision, the idea of participant – spectator model is useful in geometrical concept and to study the nuclear reaction mechanism. The results of our systematic studies on projectile fragmentation in the interaction of $^{84}$Kr with the emulsion's targets at around 1 GeV per nucleon and properties of alpha fragment emission from projectile are presented in this article. These results are compared with different type of projectile fragmentation at nearly the same as well as different energies. It is found that the multiplicity distribution of alpha fragments can be well represented by the Koba–Nielsen–Olesen (KNO) scaling hypothesis, and the production rate of alpha fragments is independent of the beam energy, but it increases with the increase of projectile mass.

We report a study of alpha fragments emission in the projectile fragmentation using $^{84}$Kr-beam as projectile having initial (energy at the point of entrance into the detector) kinetic energy of around 1 A GeV interactions with emulsion-detector's targets. The aim of the present work is to perform systematic studies on alpha fragments emission in projectile fragmentation. Projectile fragments have considerable advantages, as compared with classical experiments on the disintegration of target nuclei. Particularly, there is no threshold for the detection of projectile fragmentation products and they can be reliably identified and easily distinguished by the detector used in the present experiment. This article also presents a detailed description on mean free path, cross-section and target identification in nuclear emulsion experiments. The experimental details and measurement techniques with projectile's mean-free-path are summarized in Section 2. Target separation is crucial in emulsion-detector. Therefore, a better heavily-ionizing charged particle distribution with high statistics is presented in Section 3 for making clear-cut selection criteria of targets. Section 4 shows results on the alpha fragmentation.

## II. EXPERIMENTAL DETAILS

The data was collected by the scanning of NIKFI BR-2 stacks of nuclear emulsion plates having volume 10 cm x 20 cm x 0.06 cm. Plates were irradiated horizontally to around 1 A GeV $^{84}$Kr beam at the SIS synchrotron at GSI, Darmstadt in Germany. The flux intensity was ~$10^3$ particles per $cm^2$.

There are two standard methods for scanning of the emulsion plates, one is the line scanning and other is volume or/and area scanning. In the line scanning method, tracks are followed along their length till they interact with one of the photographic emulsion material or escape through any two of the surfaces of the emulsion or stopped in the plate. While in volume scanning, emulsion plates are scanned strip by strip and event information is collected.

Data used in this analysis was collected by line scanning method with the help of the Olympus BH-2, transmitted light-binocular microscope under 100X oil emersion objectives and 15X eyepieces [1]. The beam tracks were picked up at a distance of 4 mm from the edge of the plate and carefully followed by line scanning method. A total of 700 interactions of $^{84}Kr$ with the nuclei of the emulsion were observed by following a primary track length of 5250.0 cm, which led to a mean free path of $\lambda = 7.5 \pm 0.28$ cm. These results compliments the previously measured values of $^{84}Kr$ (6.76±0.21) [2] and (7.10±0.14) at same beam energy [3].

For each event the multiplicity of shower particles ($N_s$) mainly consists of freshly created or newly produced charged pions and kaons. Target and projectile associated fragments ($N_h$) and ($N_f$), respectively were determined. The secondary charged particles were classified according to their range in the emulsion and the relative ionization with a sensitivity of 28±1 grains per 100 μm for a singly charged minimum ionizing particle ($I_o$).

The shower particles are singly charged relativistic particles with a velocity $\beta \geq 0.7$ and very small ionization less than $1.4I_o$ in the emulsion with energies above 70 MeV, outside the projectile fragmentation cone and might be contaminated with small fraction of fast protons having energies above 400 MeV. In this experiment the projectile fragmentation cone defined by the critical angle, is $\theta_c \sim \pm 10^o$ at 0.90 A GeV [4]. The grey particle tracks ($N_g$) are believed to be associated with the recoiling protons of the target in the energy range of 30 – 400 MeV. The ionization of these tracks are $1.4I_o < I < 6.8I_{o.}$ The residual range (L) is greater than 3 mm in emulsion and has a velocity lies between $0.3 \leq \beta < 0.7$. The black particle tracks ($N_b$) are the fragments emitted from the excited target nuclei having ionization $I \geq 6.8I_{o.}$ This ionization value corresponds to proton with energies≤30 MeV. The residual range of the black track is less than 3 mm in emulsion and has a velocity $\beta < 0.3$. The heavily ionizing charged particles ($N_h$) is the sum of black and grey particles and are part of the target nucleus.

Projectile fragments (PF) are the spectator parts of the projectile nucleus with charge (Z) ≥ 1 and having velocity close to the beam velocity. The ionization of PF's is nearly constant over a few mm of range and emitted within a highly collimated forward narrow cone whose size depends upon the available beam energy. Singly charged projectile fragments, which have velocities nearly equal to the initial beam velocity and their specific ionization may be used directly to estimate their charge and number of such particles in an interaction denoted by $N_{z=1.}$

At relativistic energies, multiple charged fragments with charge Z≥2 emitted from the breakup of the projectile essentially travel with the same speed of the beam. These energetic projectile fragments are recorded in emulsion with 100% detection efficiency and this intrinsic feature of emulsion makes it a unique detector among all the particle detectors currently in use [5]. Doubly charged projectile fragment or alpha nucleus, number of such particles in an interaction is denoted by $N_\alpha$ are distinct from the singly charged projectile fragment because the ionization is directly proportional to $Z^2$ and the speed of all projectile fragments are supposed to be the same.

Multi-charge projectile fragments or heavy projectile fragment having charge Z ≥ 3 and the number of such particles in an interaction is denoted by $N_f$. The distinction between the projectile and the target spectator i.e. fragments is easy to make because the projectile like fragments corresponding to the spectator part are distributed in a forward narrow cone and there are no visual change in their grain density up to several mm, while the emitted particles and rescattered protons have much broader distribution. The fragments emitted from the target are observed as highly ionizing particles and isotropically distributed around the interaction vertex.

A total of 1302 alpha-projectile tracks were selected in minimum biased inelastic events. In each event, we recorded information about multiplicity, charge and angle of the fast singly charged particle ($N_p$), helium nuclei and multi-charged fragment, additionally information about black, grey and shower particles were also collected. The charge of the projectile fragments ($Z \geq 2$) were determined by the combination and separate methods of charge estimation, such as grain density, grain and blob density, gap-length coefficient, delta-rays density, along the each track of outgoing fragments and beam width measurement, residual range etc along the beam track [4].

## III. TARGET SEPARATION IN EMULSION DETECTOR

Photographic nuclear emulsion is a composite target detector. It composed of mainly several nuclei [6] such as H, C, N, O, Ag and Br. The incident projectile will interact with either one of the targets. These emulsion targets are generally classified into three major classes which are combination of Ag & Br nuclei having averaged $A_T$ = 94 for heavy; C, N and O nuclei having averaged $A_T$ = 14 for medium and the free hydrogen nucleus having $A_T$ = 1 for light targets.

The number of heavily ionizing charged particles depends upon the target break-up. Therefore, the peaks in $N_h$ distribution, must have strong correlation with the classes of emulsion target. Figure 1 is the normalized $N_h$ distribution for several projectile interactions with emulsion. We used such distribution to fix criteria for target separation. Four very distinct regions can be seen from the $N_h$-distribution and each region belongs to the known target class. The target separation was achieved by applying restrictions on the number of heavily ionizing charge particles and on residual range of black particles emitted in each event. Usually the targets are separated statistically on the basis of $N_h$ values, but there will always be a contamination of peripheral Ag/Br events in H and CNO events.

The present analysis has been done on an event by event basis, with the help of the following criteria:

**Ag / Br target events:** $N_h \geq 8$ and at least one track with R < 10 μm is present in an event. This class of target was the cleanest interaction and high statistics can make further separation between Ag and Br target interaction with high enough accuracy. It can be seen from the figure 1 that interactions having $N_h$>21 will be of the Br-target class with small fraction of Ag-target events [6].

**CNO target events:** $2 \leq N_h \leq 8$ and no tracks with R< 10 μm are present in an event. This class always contains very clean interaction of CNO target.

**H target events:** $N_h \leq 1$ and no tracks with R<10 μm are present in an event. This class include all $^{84}$Kr+H interactions but also some of the peripheral interactions with CNO and the very peripheral interactions with Ag / Br targets.

On the basis of the above criteria we obtained 13.4, 39.0 and 47.6 percent of interactions with H, CNO and Ag / Br targets, respectively as depicted in Figure 2.

In principal, the percentage of target interaction with incident projectile should depend on the projectile mass number and its energy due to the change in cross section. The target types are fixed in the emulsion detector. The dependence of target interactions with projectiles of different energies is depicted in Figure 2 for each target group. It can be seen that H-target shows weak dependence with projectile mass number while other two target groups are almost independent due to the admixture of different centrality events of other target group as shown in figure 1.

# IV. EXPERIMENTAL RESULTS

The measured value of the projectile and emulsion target cross section ($\sigma_{nucl}^{ApAt}$) in the interactions of primary $^{84}$Kr beam with emulsion nuclei at ~ 1 A GeV has been compared with those obtained in a very wide range of interactions of different projectile species (from $^{4}$He to $^{208}$Pb) having different energy, ranging from 1 to 200 A GeV with emulsion target nuclei. The dependence of the nuclear cross section of the projectile-target system on the projectile ($A_p$) and target ($A_t$) mass number are shown in Figure 3. This graph can be investigated according to the Bradt and Peter's formula [7]. We can parameterized the nuclear inelastic cross section for nucleus-nucleus (AA) at high incident energies by a simple geometrical formula

$$\sigma_{nucl}^{ApAt} = \pi r_o^2 \left(A_p^{1/3} + A_t^{1/3} - b\right)^2, \qquad (1)$$

where $r_o$ is the interaction radius and b is the impact parameter which is connected to the transparency of the nuclei and can be obtained by fitting the experimental data. The nuclear cross sections are seen to be almost independent of incident beam energy and substantially dependent on the atomic mass number of the colliding nuclei. It also shows a linear behaviour and fitted parameter $r_o$ = 1.30±0.02 fm and b = 1.17±0.07 can be used to reproduce the data.

## Emission of Z = 2 Projectile Fragments

A study of projectile and target fragmentation processes provides ample information about the nuclear structure. Projectile fragmentation at high energies has proven to be an especially powerful tool in the production and study of new exotic nuclei. When energies are very high and the incident nuclei are relatively massive, the clear cut participant – spectator picture fails and the concept of two emission sources for understanding the projectile fragmentation process appears to be appropriate. In order to select the model which best reproduced the experimental situation, it is necessary to carry out a detailed study of various properties of the interactions produced in different projectile-target combinations at relativistic energies. The normalized multiplicity distribution of alpha emitted from $^{84}$Kr interactions with emulsion targets at around 1 GeV per nucleon is presented in figure 4 along with several different projectiles of different energies. Almost all projectiles have nearly similar incident energy except $^{197}$Au and $^{208}$Pb having energy 10.6 and 160 GeV per nucleon, respectively. It can be seen that the multiplicity distribution of alpha projectile fragments becomes broader with the increase of the projectile mass number. This is physically expected since the number of participant helium clusters becomes larger for heavier projectile. The probability of emission of zero alpha events is always higher than the rest type of events and it decreases about more than 30% as projectile mass number increases from 14 to 208 due to the increase in number of central collisions. The reason could be that here we are dealing with events where most collisions are between nearly equal sized nuclei, for which the strict geometrical criterion is not the only condition for central collision. In the case of collision between nuclei of unequal size, it is possible to completely cover the smaller nucleus by the larger one and ideal conditions for centrality can be satisfied. On the other hand, if interacting nuclei are of equal sizes then the probability for central collisions is quite small .It can also be seen from this distribution that up to $^{132}$Xe, the emission of alpha fragments follow almost similar trend but alpha emitted from low energy $^{197}$Au interactions are showing different emission probabilities for more than two helium fragments and larger number of helium fragments are showing slightly higher emission probability. Emission probability of helium fragments from high energy $^{197}$Au and $^{208}$Pb projectiles are same within error but showing low probability for less number of helium fragments. It is clear that emission probability of large number of helium fragments is almost independent from projectile mass number and energy.

In the present paper we carry out a detailed investigation of alpha projectile fragment distribution of around 1 A GeV $^{84}$Kr emulsion interactions, including the alpha projectile fragment multiplicity distribution and the dependence of average multiplicity on the projectile mass number. The different multiplicity distributions of helium fragments obtained from the different interaction of $^{84}$Kr nuclei with different target groups of emulsion nuclei are presented in figure 5. Emission probability for larger number of helium nucleus emitted in an event such as 5 and 6 have no significant dependence on target group while strong dependence is evident for the most probable events having no emission of helium nucleus. Nevertheless small differences may be evident as: after $N_\alpha = 0$, the probability of emission of $N_\alpha = 1$, 2 & 3 in case of the interactions of $^{84}$Kr with heavy target nuclei are more; the most violent higher temperature processes than in case of with light nuclei interactions due to gentle low temperature processes. The difference in the emission probability of $N_\alpha = 4$ and 5 may depend on the target and can be seen from the figure 5 and larger emission probability is related to the H target rather than to the others. From this distribution, one may conclude that the helium emission from projectile during collision is evidently influenced by the impact parameter of the collision.

Table 1 presents the partial and the total nuclear cross section of various channels of alpha fragments ($\sigma_{N\alpha}$) emission of $^{84}$Kr interactions with emulsion as well as different emulsion target groups as mentioned in section 1 in comparison with the corresponding values of other colliding symmetric and non-symmetric systems at different energies. It can be seen from the table 1 that the total and partial nuclear cross sections of alpha fragment emission channels for $^{84}$Kr (only $^{84}$Kr data is presented in table 1) and other $^{24}$Mg [4], $^{28}$Si [4] and $^{32}$S [8] primary beams incident on emulsion nuclei result to be the same at different energies, within experimental error [4, 5, 9]. The cross section of emission of single and double alpha fragments strongly depend on the mass of the target. It can also be seen from the table 1 that the production cross sections of alpha fragments seems to be energy independent but increasing with the increase of the mass number of both projectile and target.

Emission of alpha fragment is well studied and has been proven that it is a surface phenomenon according to the participant-spectator model [10] because spectator part of projectile is mainly responsible for alpha fragments emission at just relativistic projectile energies. At high energy, only spectators of projectile and target could be related to liquid-gas phase transition. Most of the violent process could be occurred in the participant region and therefore alpha fragments emission from this region is least probable. Emission of maximum numbers of alpha fragments in an event is strongly correlated to the average multiplicity of that particle as shown in Figure 6a and the empirical relation is $N_{\alpha\,(max)} = (0.39\pm0.75)*\langle N_\alpha\rangle + (3.24\pm0.31)$.

The average multiplicity of alpha fragments produced in the final state of the reaction has strong correlation with mass number of the projectile involve in the initial state of the reaction. To see the correlation between maximum numbers of alpha fragments emission in an event as a function of projectile mass number is presented in figure 6b. It can be seen from figure 6b that the maximum numbers of alpha fragments grows as $A^{2/3}$ and the empirical relation is $N_{\alpha\,(max)} = (0.30\pm0.04)*A^{2/3} + (2.08\pm0.58)$. It can also be seen from both the figures that alpha fragments emission is a nuclear surface phenomenon and has two source emission.

The multiplicity distributions of the produced fragments have been regarded as a potentially useful source of informations of the underlying production mechanism. The validity of the KNO [20] scaling of the alpha fragments produced via the decay properties of the excited projectile in high energy heavy ion interactions using the two-source emission picture has been confirmed by experimental group [21]. The KNO scaling hypothesis was originally derived by assuming the Feynman scaling of the inclusive particles production cross section. This scaling is a consequence of the nuclear geometry, which is energy independent and multiplicities of produced alpha particles, $P(N_\alpha)$ in high energy interactions obeys the scaling law,

$$\Psi(Z) = 4Z \exp(-2Z), \qquad (2)$$

with

$$\Psi(Z) = \langle N_\alpha \rangle P(N_\alpha) = \langle N_\alpha \rangle\, \sigma_{N\alpha} / \sigma_{nucl}. \qquad (3)$$

Here $\Psi(Z)$ should be an energy independent function of the scaled variable, $Z = N_\alpha/\langle N_\alpha \rangle$ is the number of alpha fragments produced in an event normalized by the average number of alpha fragment of the whole data set, $P(N_\alpha)$ is the probability of finding $N_\alpha$ fragments in the final state of interactions, $\sigma_{N\alpha}$ is the total cross section for the specific reaction channel with a state of multiplicity of $N_\alpha$ and $\sigma_{nucl}$ refers to the total nuclear inelastic cross section. The multiplicity of the produced alpha fragments from the events of different projectiles over a wide range of energies can be represented by a universal experimental function of the following form [21]:

$$\Psi(Z) = AZ \exp(BZ), \qquad (4)$$

where A and B are constants, whose values are different in different literature. Therefore, best fit function method has been adopted to determine a accurate and unique value for each of A and B, which is the best fit value for all experimental points.

In figure 7, we plotted $\langle N_\alpha \rangle P(N_\alpha)$ as function of the scaled variable $N_\alpha/\langle N_\alpha \rangle$ for alpha fragments emission in $^{84}$Kr interactions with different target group of emulsion at around 1 A GeV. Experimental data are fitted with equations (2) and (4). It can be seen from this figure that almost all data points fall on the same universal curve. The fitting parameters A and B are 4.14±0.46 and 2.21±0.07, respectively. Within statistical errors, these values are close to the theoretical values of A=4 and B=2, and there is no significant difference evident for small number of alpha fragments emission but for large number of alpha fragments emission significant difference is observed.

Figure 8 presents the multiplicity distribution of $<N_\alpha>P(N_\alpha)$ as a function of the scaled variable $N_\alpha/<N_\alpha>$ for the alpha fragments considered through all this work, which are then compared also to the universal KNO scaling. Experimental data points are presented by the different symbols for different colliding system's having projectile energies ranging from 0.95 to 200 A GeV, while the solid and dashed curves are the results of equations (3) and (4), respectively. There fitted parameters values for A and B are 5.10±0.11 and 2.23±0.07, respectively. The upper and lower insets are presenting similar distributions for high and low energy experiments. The fitted parameters for the same are 3.18±0.42, 2.07±0.08 and 3.98±0.39, 2.26±0.04, respectively. From the above values and figure we can seen that the fitted parameter's values are almost similar that reflects true sense of the KNO scaling.

If multiplicity scaling is valid, as a consequence also the moments defined in Ref. [21]

$$C_q = <N^q_\alpha>/<N_\alpha>^q, \qquad (5)$$

for q = 2, 3, 4 and 5 relative to the alpha fragments, should also be energy independent. To test the validity of KNO scaling, the multiplicity distributions using the $C_q$ as given by equation (5) should be studied. Table 2 presents the results on the average multiplicities and the moments $C_2$, $C_3$, $C_4$ and $C_5$ of the alpha fragments emitted from the collision of 1 A GeV $^{84}$Kr projectile with the target emulsion nuclei (H, CNO, Ag/Br and Em) compared with the corresponding values for $^{12}C$, $^{16}O$, $^{22}Ne$, $^{28}Si$, $^{32}S$, $^{197}Au$ and $^{208}Pb$ beams at various energies. The second Muller moment [22] can be calculated from the following equation:

$$F_2 = (C_2 - 1) \langle N_\alpha\rangle^2 - \langle N_\alpha\rangle, \qquad (6)$$

and it is included in table 2 for all projectiles and energies considered here. It is clear that the values of $C_2$ and $C_3$ moments, within the experimental errors, do not seem to depend upon the energy or masses of the colliding nuclei. On the other hand, the higher moments $C_3$, $C_4$ and $C_5$ show a slow increase in their values as the projectile mass number increases. The second Muller moments are non-zero: this may indicate a strong correlation among the alpha fragments. In addition to that, the mean multiplicity $\langle N_\alpha\rangle$ derived in the interactions of different projectiles at various incident energies can be satisfactory described in terms of the projectile mass number $A_p$ by the following power law:

$$\langle N_\alpha\rangle = {}_cA^d{}_p, \qquad (7)$$

where c = 0.37±0.02 and d = 0.47±0.01. An interesting observation of these experiments is that the value of the ratio $\langle N_\alpha\rangle/D$ ($D=\langle N^2{}_\alpha\rangle-\langle N_\alpha\rangle^2$) for all projectiles is approximately equal to the constant, revealing that asymptotic multiplicity scaling and is equal to that observed in hadron-nucleus interactions [21]. These exhibit an almost identical behavior in all beams at different energies, which leads to an energy and masses of colliding nuclei independent effect mechanism on the breakup of projectile nuclei through alpha fragments [21].

The mean multiplicities of the different charged secondary particles emitted from the participant and target spectator regions of the interactions of around 1 A GeV $^{84}$Kr with the different emulsion target groups at different degrees of projectile nucleus disintegration are presented in table 3 with heavy projectile fragment(s) in an event. The dependence of the average multiplicities of these secondaries on the target mass is also tabulated in table 3 for the interactions of $^{84}$Kr projectile with H, CNO and Ag / Br groups of target nuclei. It can see that the averages number of multiplicities of different secondaries increase substantially with increasing target mass number except number of shower particles. Similar trends can be seen even for the different alpha fragments emission channels. It is evident that the multiplicities of all types of charged particles depend strongly on the impact parameter of nucleus-nucleus collisions [4]. The values of mean multiplicities of charged secondaries in the events in which the alpha fragments are accompanied by heavy projectile fragments having $Z \geq 3$ seem to be almost constant with in statistical error as the alpha fragments multiplicity increases and exhibit an invariant behavior of charged secondaries in the lighter projectile frame. These events can be considered coming from more peripheral collisions, indicating that the participant parts from both projectile and target nuclei are very small and the energy transferred from the projectile to the target is nearly constant. The present measurements are seen to be in good systematic agreement with the results obtained in other experiments [4, 21, 25]. It can also be seen from the table 3 that the average numbers of slow target fragments, $\langle N_b \rangle$ emitted in this experiment and from other [4, 21, 25] emulsion experiments are very similar within experimental errors. This indicates that the fragments evaporated from the target do not seem to depend either on the energy or on the mass of the beam and in case of emulsion detector; there are no major changes in the composition of emulsion in the main target group's percentage.

## V. CONCLUSIONS

It is quite interesting to study the projectile fragmentation of heavy ions such as $^{84}$Kr having initial kinetic energy of ~1 A GeV, as some of the alpha fragmentation characteristics change with the mass of the projectiles. The idea of two sources is one of them. The results obtained from the above investigation allow us to make the following main conclusions:

A detail study has been done for the separation of different target events up to the level of Ag and Br, and a clear cut cut-off value for each target group has been fixed. The total and partial nuclear cross sections of alpha fragmentation channels in relativistic and ultra-relativistic nucleus-nucleus collisions are energy independent. The production rates of the alpha fragments in very heavy projectiles such as $^{197}$Au and $^{208}$Pb are much broader and maximum number of alpha emitted up to 15. Within the statistical errors, yield of alpha fragments in the interaction of projectile with different target groups of emulsion shows no significant dependence. That is the emission of alpha fragments from the projectile is free from the influence of target. Our observations confirm the idea that, while considering the nuclear collisions induced by massive beams, the relativistic alpha fragments must be ascribed to two emission sources. One of them is the projectile itself and the other is a fireball which is formed when nucleons are mutually swept away from the projectile and the target. Based on the two-source emission picture, a kind of KNO scaling is obtained and describes the multiplicity distribution of alpha projectile fragments. It is interesting to note that the multiplicity distributions of alpha projectile fragments emitted in the interactions of different projectile with different target emulsion nuclei at different energies are well described by the KNO scaling presentation. To validate the KNO scaling the $C_q$ moments should be energy independent. The derived values of the moments have been checked. Within the experimental errors, the values of $C_2$ and $C_3$ moments do not seem to depend upon the energy and/or masses of the colliding nuclei while the higher moments $C_3$, $C_4$ and $C_5$ shows slow increase in their values as the projectile mass number increases. In our experiment, we derived non-zero second Muller moments which indicate existence of a strong correlation among the alpha fragments. The averages multiplicities of all types of charged particles depend strongly on the impact parameter of nucleus-nucleus collisions. The average values of all charged secondaries of those events in which the alpha fragments have been accompanied by heavy fragments exhibit a behavior independent of the alpha fragment multiplicity. Actually, such events are related to more peripheral collisions, where the participant parts of the projectile and target are supposed to be very small and the energy transfer between the two colliding nuclei is almost constant.

## Acknowledgements

We thank the accelerator staff at GSI/SIS and JINR (Dubna) for their help with the exposures.

# References

[1] http://www.alanwood.net/downloads/olympus-bhm-bh-2brochure.pdf

[2] V. Singh, Ph.D. Thesis, Banaras Hindu University, Varanasi (1998).

[3] S. A. Krasnov et al., Czech. J. Phys. **46**, 531 (1996).

[4] M. A. Jilany, Nucl. Phys. **A705**, 477 (2002).

[5] G. Singh, P. L. Jain, Phys. Rev. **C54**, 3185 (1996).

[6] V. Singh et al., arxiv:nucl-ex/0412051v1 (2004).

[7] H. L. Bradt and B. Peters, Phys. Rev. **77**, 54 (1950).

[8] G. Singh, A. Z. M. Ismail, P. L. Jain, Phys. Rev. **C43**, 2417 (1991); G. Singh, K. Sengupta, P. L. Jain, Phys. Rev. **C42**, 1757(1990); G. Singh, K. Sengupta, P. L. Jain, Phys. Lett. **B222**, 301 (1989).

[9] M. I. Adamovich et al., Eur. Phys. J. **A5**, 429 (1999).

[10] J. Hufner, J. Knoll, Nucl. Phys. **A290**, 460 (1977).

[11] M. El-Nadi et al., Proceedings of the 27th International Cosmic ray Conference, Hamburg, Germany, 1366 (2001).

[12] M. I. Adamovich et al., Sov. J. Nucl. Phys. **29**, 52 (1979).

[13] M. I. Adamovich et al., Phys. Rev. Lett. **62**, 2801 (1989).

[14] M. A. Jilany, Euro. Phys. J. **A 22**, 471 (2004).

[15] R. R. Joseph et al., J Phys. G : Nucl. Part. Phys. **18**, 1817 (1992).

[16] V. E. Dudkin et al., Nucl. Phys. **A509**, 783 (1990).

[17] San-Hong Fan, Fu-Hu Liu, Radiation measurements **43**, S239 (2008).

[18] A. Gill et al., Int. J. of Mod. Phys. **A5**, 755 (1990).

[19] C. J. Waddington and P. S. Freier, Phys. Rev. **C31**, 888 (1985).

[20] Z. Koba, D. H. Nielsen and P. Olesen, Nucl. Phys. **B40**, 317 (1972).

[21] M. El-Nadi et al., J. Phys. **G28**, 1251 (2002); S. Kamel, Nuovo Cimento

**A112**,733 (1999); Fu-Hu Liu et al., Cimento **A111**, 1219 (1998); Fu-Hu Liu Phys. Rev. **C62**, 024613 (2000); M. El-Nadi et al., Int. J. Mod. Phys. **E2**, 381 (1993); P. L. Jain et al., Phys. Rev. **C33**, 1970 (1986); L. S. Liu et al., Phys. Rev. **D27**, 2640 (1983).

[22] A. H. Muller, Phys. Rev. **D4**, 150 (1971).

[23] M. El-Nadi et al., Int. J. Mod. Phys. **E2**, 381 (1993).

[24] B. K. Singh et al., Nucl. Phys. **A570**, 819 (1994); M. I. Adamovich et al., Z. Phys. **A351**, 311 (1995).

[25] M. El-Nadi et al., J. Phys. **G28**, 1251 (2002).

[26] Zhang Dong-Hai et al., Chinese Phys. **B Vol. 18, No. 2**, 0531 (2009).

**TABEL I:** Different emission channels of projectile fragmentations.

| **Projectile + Target** | **$^{84}$Kr + H** | **$^{84}$Kr + CNO** | **$^{84}$Kr + Ag(Br)** | **$^{84}$Kr + Em** |
|---|---|---|---|---|
| Energy (A GeV) | 1.0 | 1.0 | 1.0 | 1.0 |
| $\sigma_{1\alpha}$ (mb) | 158±7 | 209±9 | 419±19 | 262±11 |
| $\sigma_{2\alpha}$ (mb) | 106±5 | 116±5 | 140±6 | 117±5 |
| $\sigma_{3\alpha}$ (mb) | 79±4 | 70±3 | 116±5 | 88±4 |
| $\sigma_{4\alpha}$ (mb) | 65±3 | 56±2 | 98±4 | 73±3 |
| $\sigma_{5\alpha}$ (mb) | 41±2 | 22±1 | 65±3 | 43±2 |
| $\sigma_{6\alpha}$ (mb) | 14±0.63 | 18±0.8 | 18±0.8 | 17±0.7 |
| $\sigma_{7\alpha}$ (mb) | 11±0.50 | 11±0.5 | 9±0.4 | 10±0.5 |

**TABLE II:** The average multiplicity $\langle N_\alpha \rangle$, the $C_q$ moments and the second Muller moment $F_2$ for alpha fragments emitted from projectile in different nucleus – nucleus collisions at various energies. Values are arranged according to increasing number of atomic mass number of projectiles.

| Reaction | Energy (A GeV) | $\langle N_\alpha \rangle$ | $C_2$ | $C_3$ | $C_4$ | $C_5$ | D | $F_2$ | $\langle N_\alpha \rangle$/D | Ref. |
|---|---|---|---|---|---|---|---|---|---|---|
| **$^{12}$C-Em** | 3.7 | 1.49± 0.13 | 1.20± 0.10 | 1.71± 0.15 | 2.72± 0.23 | 4.68± 0.40 | 0.67± 0.18 | -1.05± 0.23 | 2.22± 0.19 | 23 |
| **$^{16}$O-Em** | 3.7 | 1.60± 0.05 | 1.23± 0.07 | 1.81± 0.16 | 2.99± 0.35 | 5.38± 0.78 | 0.77± 0.12 | -1.01± 0.18 | 2.07± 0.41 | 8 |
| **$^{22}$Ne-Em** | 3.7 | 1.60± 0.03 | 1.29± 0.03 | 2.09± 0.05 | 3.99± 0.09 | 8.47± 0.18 | 0.86± 0.05 | -0.86± 0.08 | 1.85± 0.04 | 23 |
| **$^{24}$Mg-Em** | 3.7 | 1.71± 0.04 | 1.32± 0.03 | 2.24± 0.06 | 4.50± 0.11 | 10.11± 0.25 | 0.97± 0.05 | -0.77± 0.09 | 1.82± 0.06 | 14 |
| **$^{28}$Si-Em** | 3.7 | 1.79± 0.04 | 1.31± 0.03 | 2.14± 0.17 | 4.19± 0.46 | 9.33± 1.37 | 1.00± 0.10 | -0.80± 0.19 | 1.83± 0.07 | 24 |
| **$^{28}$Si-Em** | 14.6 | 1.78± 0.05 | 1.34± 0.04 | 2.31± 0.07 | 4.71± 0.14 | 10.74± 0.32 | 1.09± 0.03 | 0.70± 0.02 | 1.63± 0.08 | 25 |
| **$^{32}$S-Em** | 200.0 | 1.83± 0.06 | 1.35± 0.13 | 2.35± 0.35 | 4.95± 1.04 | 11.90± 3.36 | 1.08± 0.20 | -0.66± 0.44 | 1.55± 0.07 | 8 |
| **$^{40}$Ar-Em** | 1.9 | 2.20± 0.09 | 1.34± 0.06 | 2.19± 0.10 | NA | NA | 1.28± 0.12 | -0.55± 0.29 | 1.71± 0.08 | 8 |
| **$^{56}$Fe-Em** | 1.9 | 2.53± 0.09 | 1.34± 0.05 | 2.21± 0.08 | 4.31± 0.16 | 9.57± 0.35 | 1.48± 0.12 | -0.35± 0.33 | 1.70± 0.07 | 8 |
| **$^{84}$Kr-Em** | 1.5 | 2.91± 0.08 | 1.39± 0.04 | 2.40± 0.07 | 4.87± 0.14 | 11.14± 0.32 | 1.82± 0.12 | 0.39± 0.35 | 1.60± 0.06 | 8 |
| **$^{84}$Kr-Em** | 1.8 | 3.07± 0.03 | 1.35± 0.08 | 2.23± 0.22 | 4.29± 0.60 | 9.24± 1.73 | 1.82± 0.21 | 0.23± 0.75 | 1.68± 0.07 | 26 |
| **$^{84}$Kr-H** | 0.95 | 1.75± 0.36 | 1.46± 0.05 | 2.67± 0.06 | 4.70± 0.14 | 10.17± 0.29 | 1.68± 0.08 | -0.34± 0.17 | 1.04± 0.06 | **Present work** |
| **$^{84}$Kr-CNO** | 0.95 | 1.53± 0.14 | 1.34± 0.03 | 2.60± 0.05 | 4.69± 0.12 | 10.43± 0.30 | 1.50± 0.06 | -0.73± 0.15 | 1.02± 0.03 | **Present work** |
| **$^{84}$Kr-Ag(Br)** | 0.95 | 2.81± 0.29 | 1.43± 0.04 | 2.66± 0.05 | 4.58± 0.11 | 10.70± 0.28 | 1.62± 0.08 | 0.59± 0.18 | 1.73± 0.07 | **Present work** |
| **$^{84}$Kr-Em** | 0.95 | 2.63± 0.06 | 1.48± 0.04 | 2.64± 0.06 | 5.33± 0.13 | 10.83± 0.32 | 1.82± 0.10 | 0.69± 0.10 | 1.45± 0.02 | **Present work** |
| **$^{197}$Au-Em** | 10.6 | 4.72± 0.24 | 1.35± 0.07 | 2.21± 0.11 | 4.13± 0.21 | 8.58± 0.44 | 7.83± 0.17 | 3.08± 0.16 | 0.60± 0.05 | 9 |
| **$^{208}$Pb-Em** | 160 | 4.47± 0.26 | 1.37± 0.08 | 2.22± 0.13 | 4.13± 0.24 | 8.44± 0.49 | 7.13± 0.18 | 2.92± 0.17 | 0.63± 0.06 | 5 |

**TABLE III:** Mean multiplicity values of other secondary particles related to the participants and target spectator regions for different alpha fragments emission channels in different emulsion target groups.

| Reaction | Multiplicities | 1α | 2α | 3α | 4α | All α's |
|---|---|---|---|---|---|---|
| $^{84}$Kr-H | $<N_s>$ | 06.14± 0.55 | 11.39± 1.03 | 09.0± 0.81 | 11.00± 0.99 | 08.45± 1.09 |
| | $<N_g>$ | 00.29± 0.02 | 00.28± 0.02 | --- | 00.29± 0.02 | 00.28± 0.03 |
| | $<N_b>$ | 00.29± 0.02 | 00.22±0.01 | 00.14± 0.01 | 00.43± 0.03 | 00.25± 0.03 |
| | $<N_h>$ | 00.57± 0.05 | 00.50± 0.04 | 00.14± 0.01 | 00.72± 0.06 | 00.53± 0.07 |
| $^{84}$Kr-CNO | $<N_s>$ | 07.50± 0.75 | 08.50± 0.76 | 06.06± 0.55 | 09.25± 0.83 | 05.81± 0.52 |
| | $<N_g>$ | 02.50± 0.25 | 02.29± 0.21 | 02.41± 0.22 | 04.75± 0.43 | 01.94± 0.17 |
| | $<N_b>$ | 02.62± 0.26 | 03.18± 0.29 | 02.41± 0.22 | 04.25± 0.38 | 02.49± 0.23 |
| | $<N_h>$ | 05.12± 0.51 | 05.47± 0.49 | 04.81± 0.43 | 09.00± 0.81 | 04.43± 0.39 |
| $^{84}$Kr-AgBr | $<N_s>$ | 13.07± 1.96 | 13.05± 1.69 | 15.49± 2.16 | 18.00± 3.60 | 14.10± 0.99 |
| | $<N_g>$ | 06.57± 0.98 | 06.95± 0.90 | 08.71± 1.21 | 07.63± 1.52 | 07.33± 0.52 |
| | $<N_b>$ | 06.91± 1.03 | 07.92± 1.02 | 08.16± 1.14 | 08.13± 1.62 | 07.82± 0.55 |
| | $<N_h>$ | 13.48± 2.02 | 14.97± 1.94 | 16.87± 2.36 | 15.76± 5.12 | 15.15± 1.06 |
| $^{84}$Kr-Em | $<N_s>$ | 10.56± 1.26 | 11.55± 1.03 | 12.51± 1.50 | 15.60± 2.49 | 10.52± 0.53 |
| | $<N_g>$ | 04.62± 0.55 | 04.56± 0.41 | 06.28± 0.75 | 05.83± 0.93 | 04.53± 0.23 |
| | $<N_b>$ | 04.88± 0.58 | 05.33± 0.47 | 05.39± 0.64 | 06.14± 0.98 | 04.97± 0.25 |
| | $<N_h>$ | 09.50± 1.14 | 09.89± 0.89 | 11.67± 1.40 | 11.97± 1.91 | 09.50± 0.47 |

## FIGURE CAPTIONS

**FIG.1:** Normalized heavily ionizing charged particle multiplicity distribution.

**FIG 2:** Percentage of target interactions as a function of projectile mass number for fixed target emulsion experiments.

**FIG. 3:** The square root of the nuclear cross section as a function of ($A_p^{1/3}+A_t^{1/3}$) are plotted for $^{84}$Kr interactions together with data obtained with different projectiles (from $^{4}$He to $^{208}$Pb) of different energies (from 1 to 200 A GeV) on nuclear emulsion nuclei.

**FIG. 4:** The normalized multiplicity distributions of alpha projectile fragments of various projectiles with emulsion interactions at different incident energies. Solid and dotted lines are the Gaussian fits on data points but thin solid line on low energy $^{197}$Au data point is just to guide the eyes.

**FIG. 5:** Multiplicity of alpha projectile fragments emitted in the interaction of $^{84}$Kr nuclei with different emulsion target groups at around 1 GeV per nucleon.

**FIG. 6:** Linear correlation between the maximum numbers of alpha fragments emitted during collisions of different mass number projectiles **(a)** versus average number of helium nucleus and **(b)** versus two third power of projectile mass number. Solid line is the best-fit. Error bar is 1 unit for all data points in case of maximum helium nucleus emission. Data points are from $^{4}$He at 3.7 [11], $^{12}$C at 3.7 [12], $^{16}$O at 2.0 [13], $^{22}$Ne at 3.5 [4], $^{24}$Mg at 3.7 [14], $^{28}$Si at 3.7 [4], $^{32}$S at 2.0 [8], $^{40}$Ar at 1.8 [15], $^{56}$Fe at 1.8 [16], $^{84}$Kr at 1.7 [17], $^{84}$Kr at 0.95 [**Present work**], $^{139}$La at 0.99 [18] and $^{197}$Au at 0.99 [19].

**FIG. 7:** $\langle N_\alpha \rangle P(N_\alpha)$ distribution as a function of the scaled variable $N_\alpha/\langle N_\alpha \rangle$ for alpha fragments emitted in the $^{84}$Kr interactions with the different emulsion target groups at around 1 A GeV energy and compared with the universal KNO scaling. Symbols are the experimental data points while the solid and dotted curves are the results of equations (2) and (4), respectively.

**FIG. 8:** $\langle N_\alpha \rangle P(N_\alpha)$ distribution as a function of the scaled variable $N_\alpha/\langle N_\alpha \rangle$ for helium fragments for different projectiles incident on emulsion target nuclei at different energies compared with the universal KNO scaling. Symbols are the data points while the solid and dotted curves are the results of equations (2) and (4), respectively. Upper and lower insets are the same distribution for higher and lower energies projectiles.

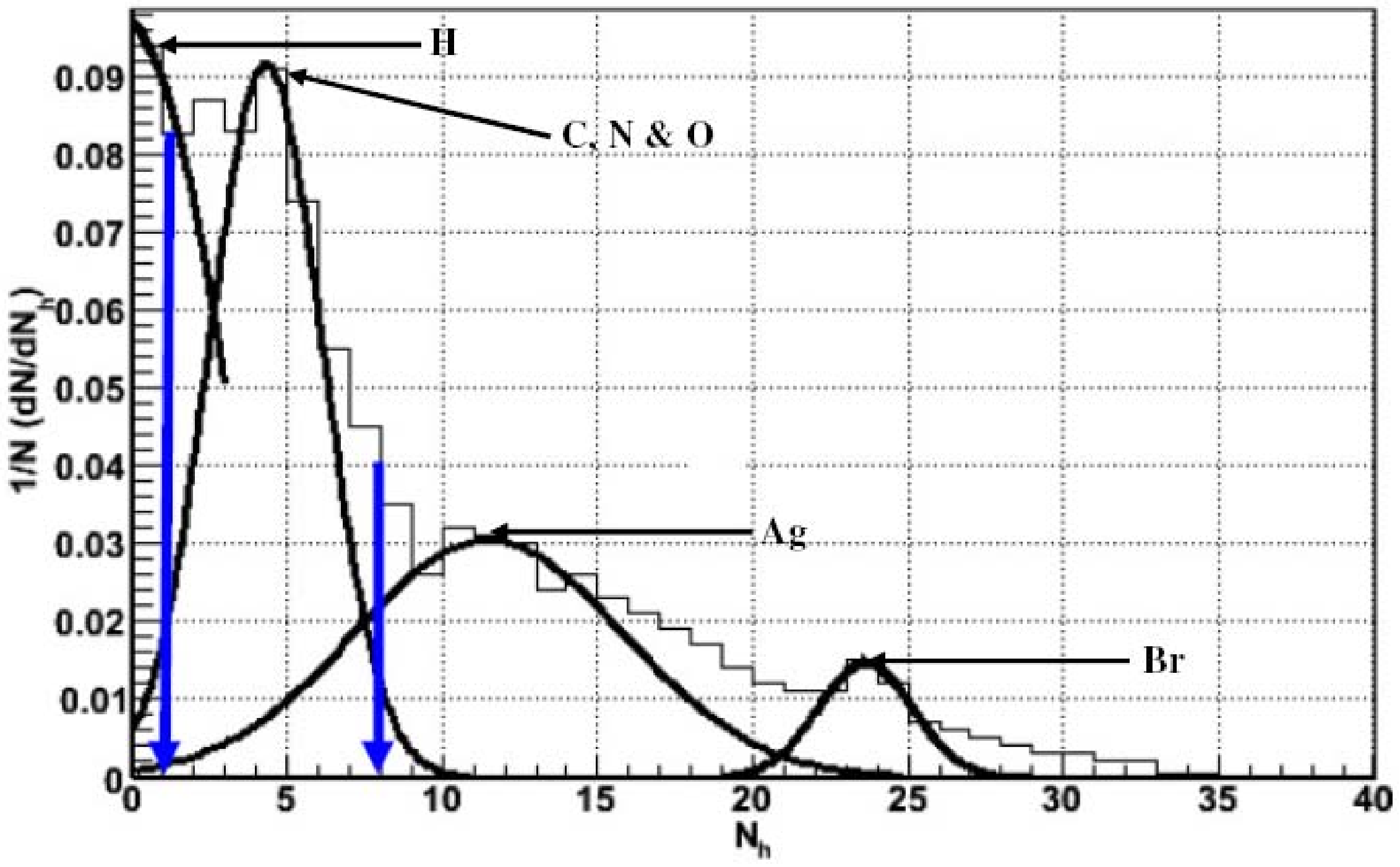

**FIG. 1**

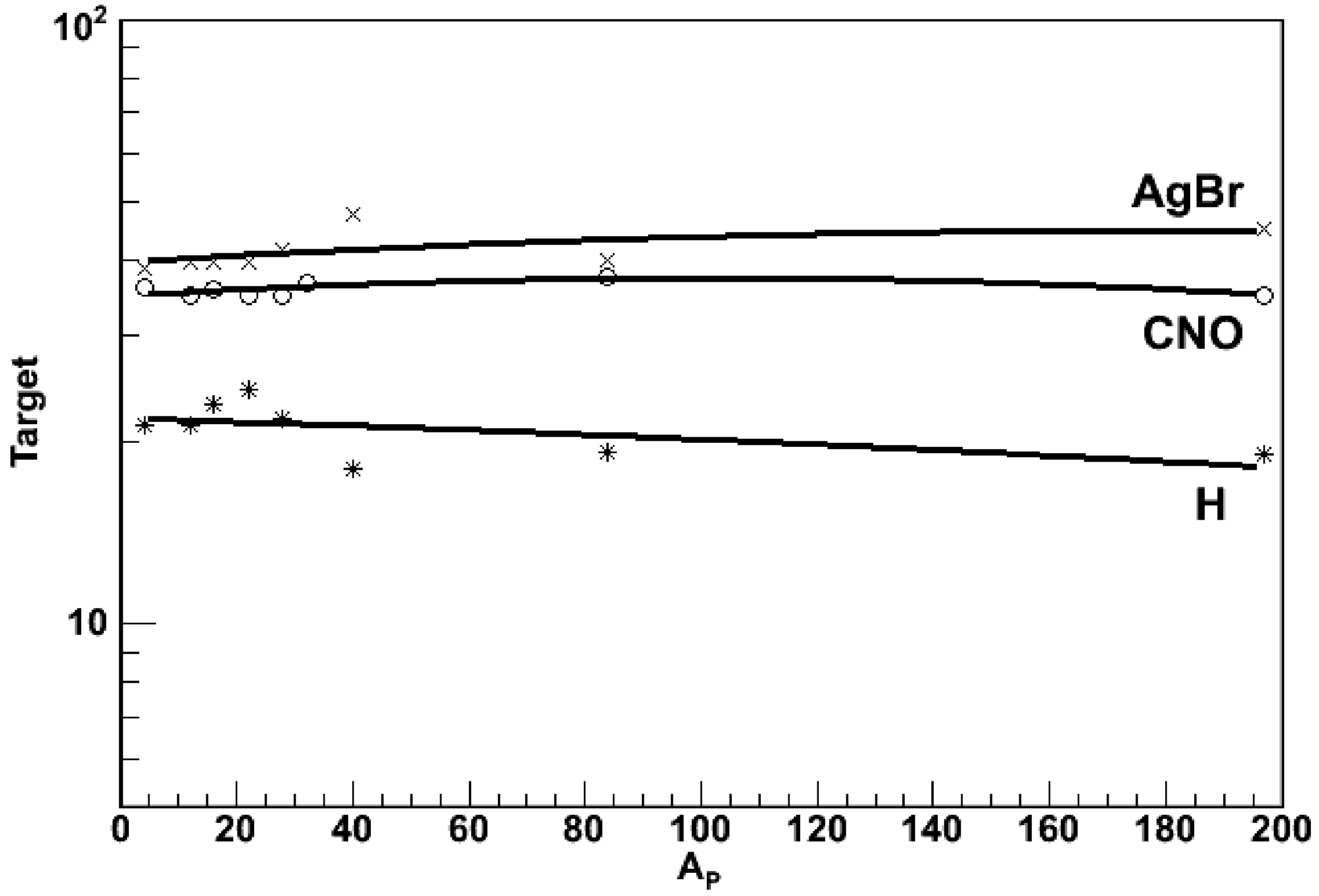

**FIG. 2:**

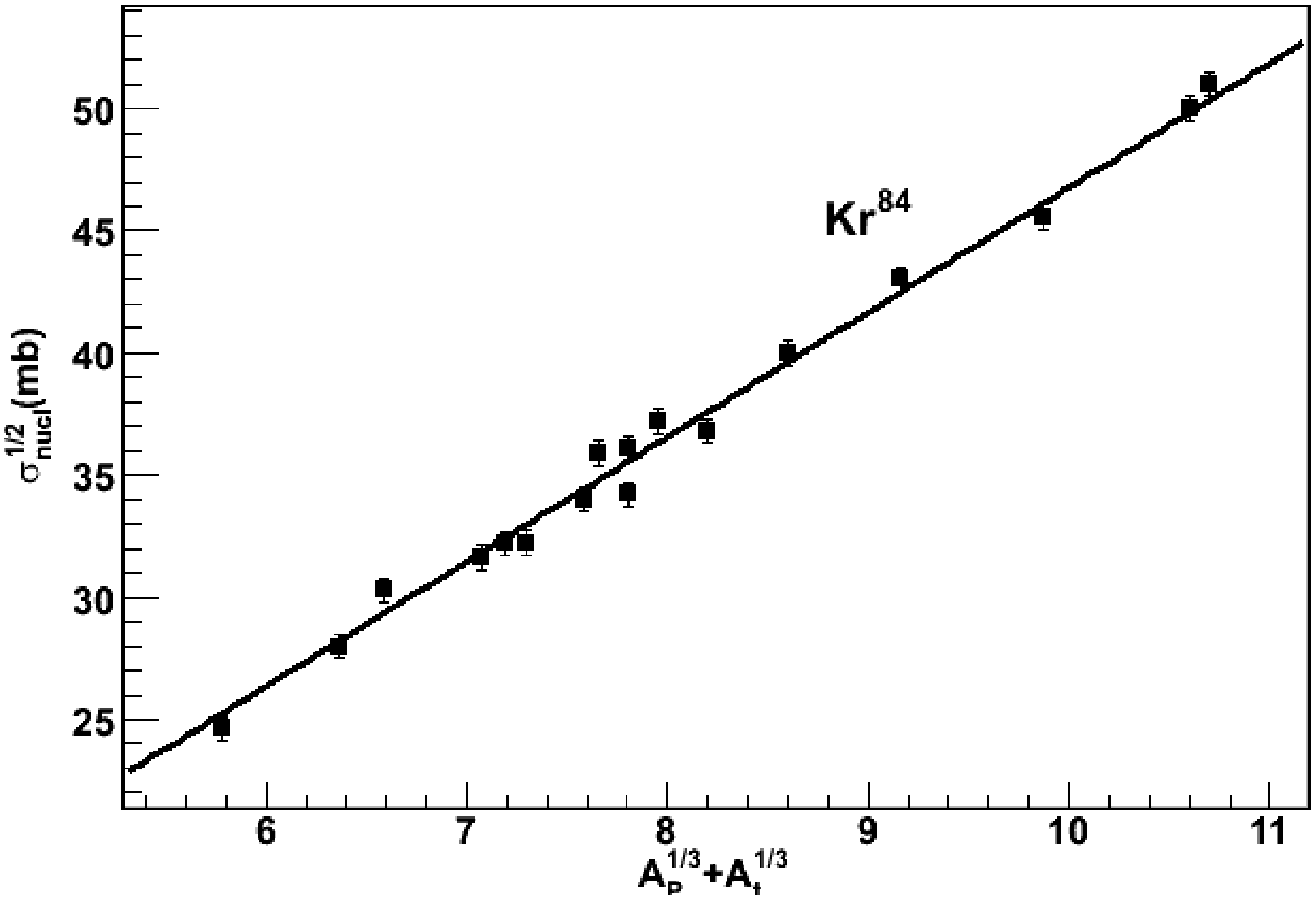

**FIG. 3:**

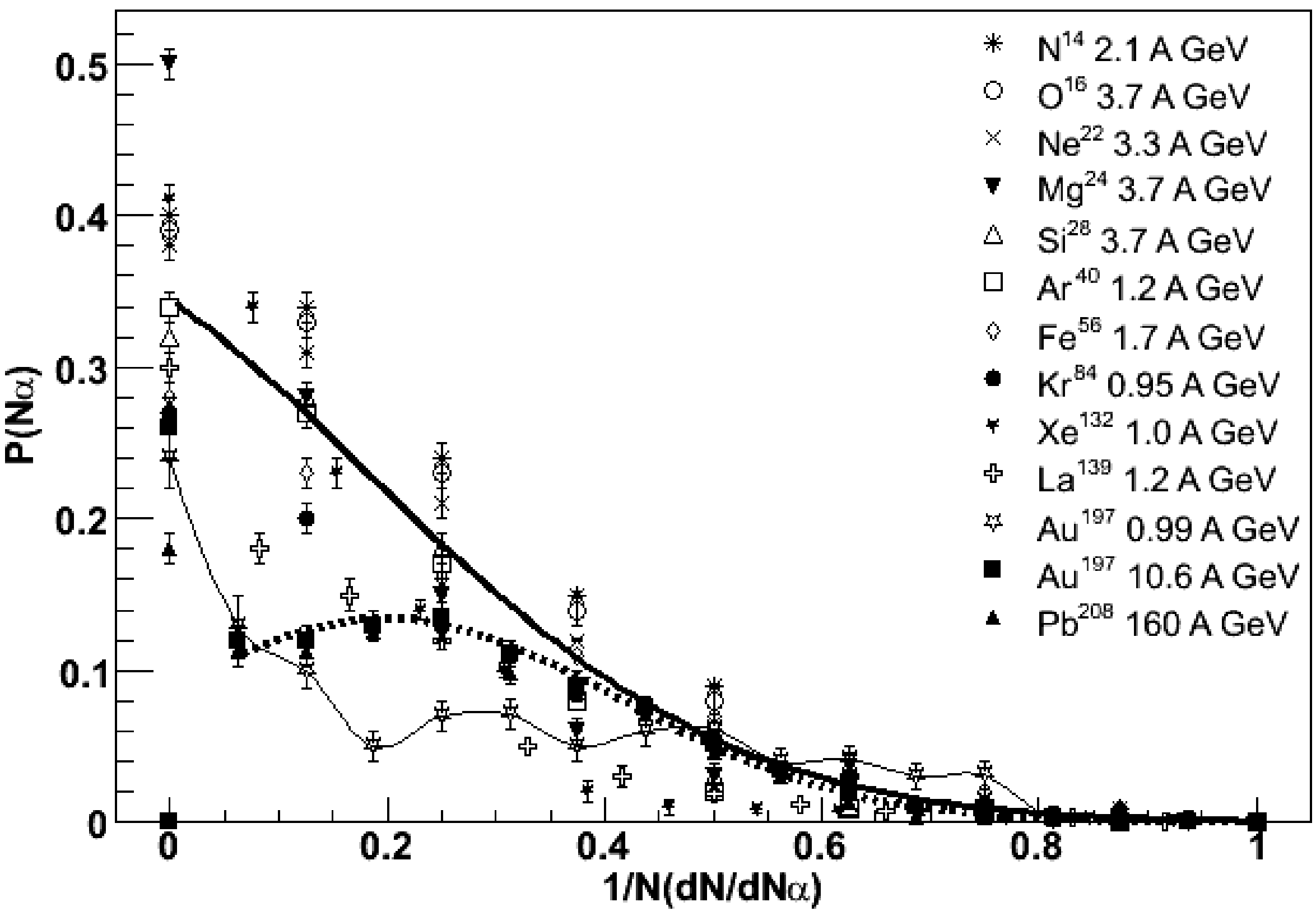

P(Nα)
1/N(dN/dNα)
0
0.1
0.2
0.3
0.4
0.5
0
0.2
0.4
0.6
0.8
1
N14 2.1 A GeV
O16 3.7 A GeV
Ne22 3.3 A GeV
Mg24 3.7 A GeV
Si28 3.7 A GeV
Ar40 1.2 A GeV
Fe56 1.7 A GeV
Kr84 0.95 A GeV
Xe132 1.0 A GeV
La139 1.2 A GeV
Au197 0.99 A GeV
Au197 10.6 A GeV
Pb208 160 A GeV

**FIG. 4:**

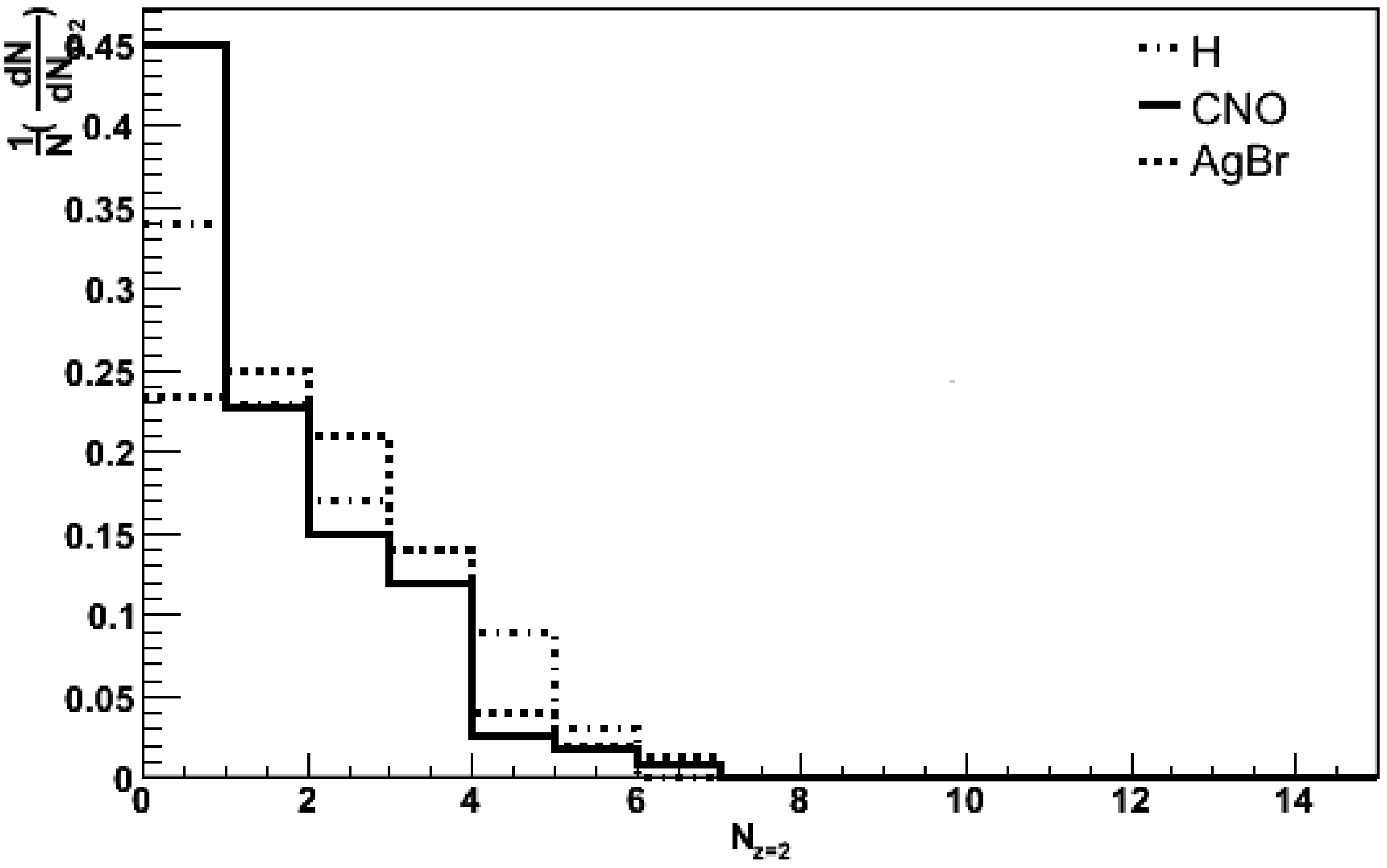

**FIG. 5:**

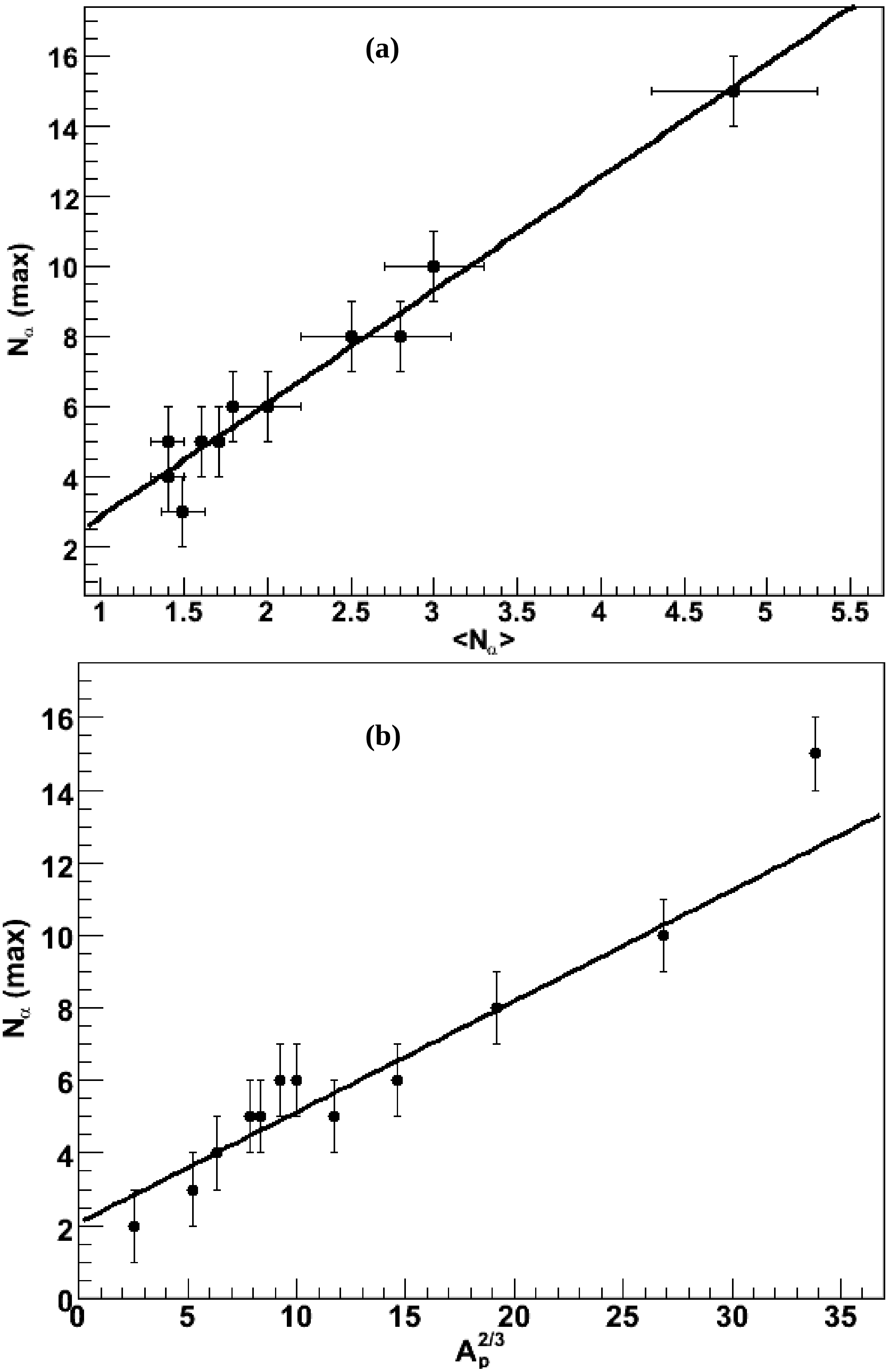

**FIG. 6:**

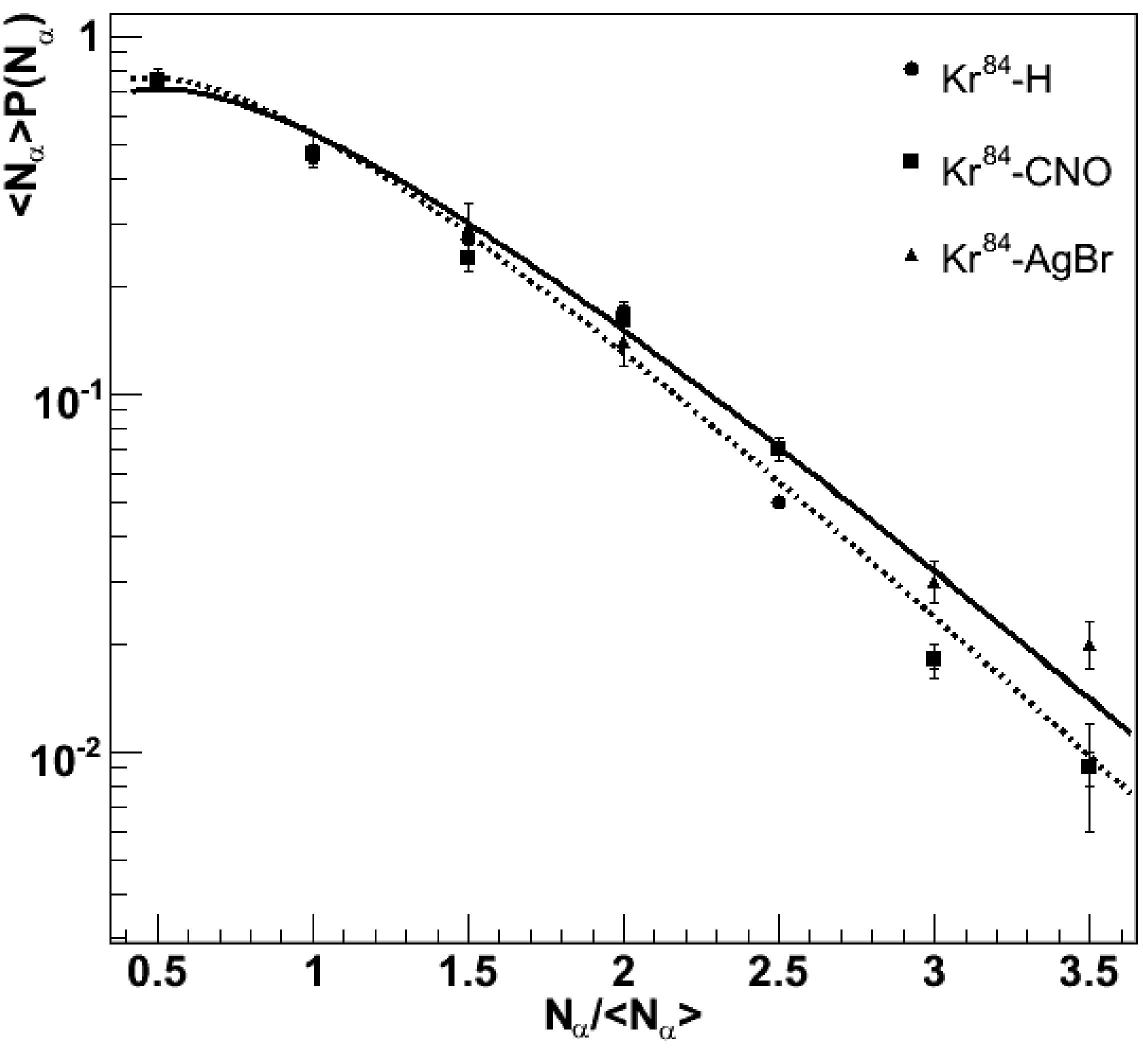

**FIG. 7:**

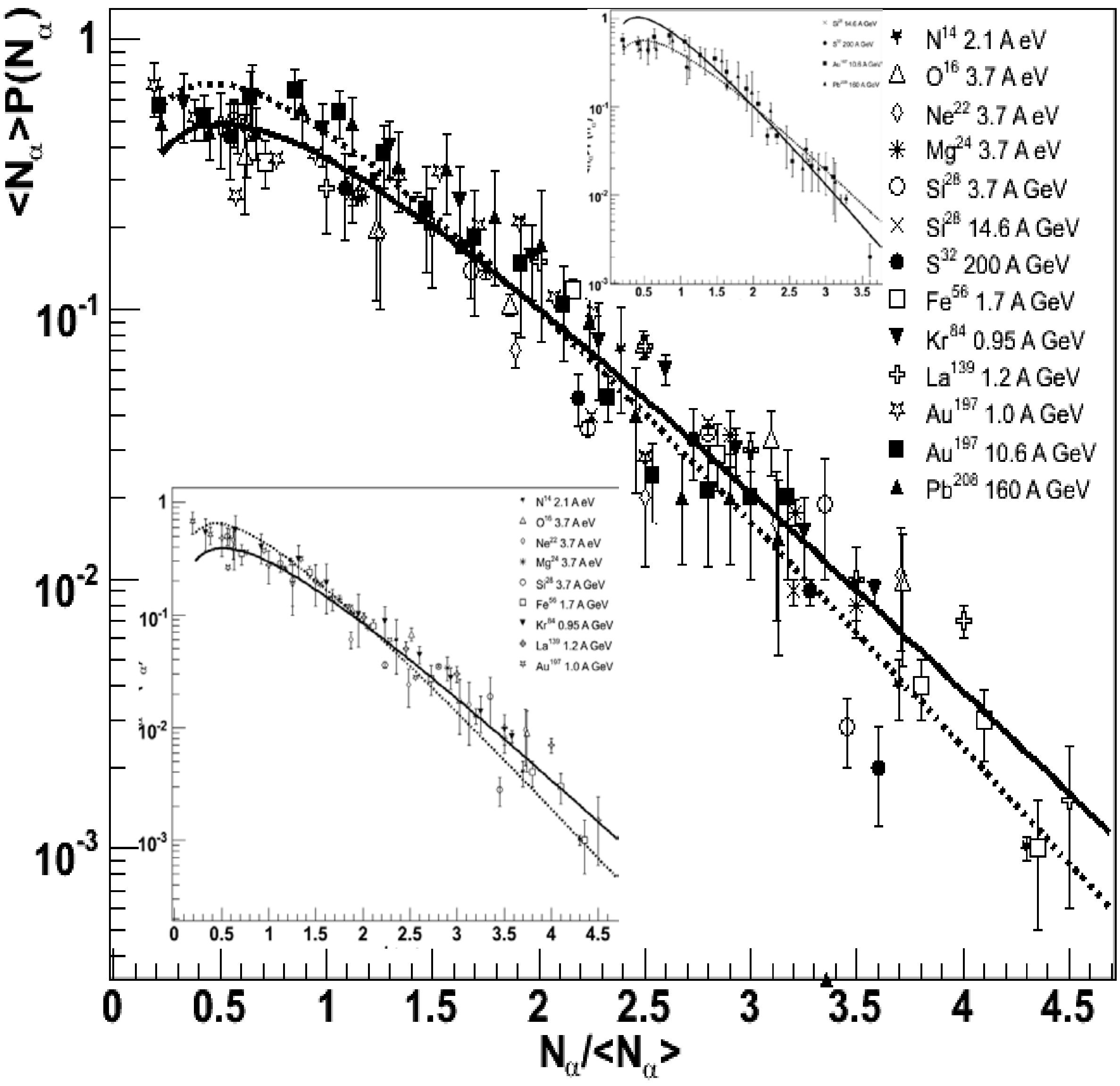

<N_α>P(N_α)
N_α/<N_α>
1
10^-1
10^-2
10^-3
0
0.5
1
1.5
2
2.5
3
3.5
4
4.5
N^14 2.1 A eV
O^16 3.7 A eV
Ne^22 3.7 A eV
Mg^24 3.7 A eV
Si^28 3.7 A GeV
Si^28 14.6 A GeV
S^32 200 A GeV
Fe^56 1.7 A GeV
Kr^84 0.95 A GeV
La^139 1.2 A GeV
Au^197 1.0 A GeV
Au^197 10.6 A GeV
Pb^208 160 A GeV

**FIG. 8:**